# A Distributed Trust and Reputation Framework for Mobile Ad Hoc Networks

Jaydip Sen

Innovation Lab, Tata Consultancy Services Ltd.,
Bengal Intelligent Park, Salt Lake Electronics Complex, Kolkata – 700091, India
`Jaydip.Sen@tcs.com`

**Abstract.** In a multi-hop *mobile ad hoc network* (MANET), mobile nodes cooperate to form a network without using any infrastructure such as access points or base stations. The mobility of the nodes and the fundamentally limited capacity of the wireless medium, together with wireless transmission effects such as attenuation, multi-path propagation, and interference combine to create significant challenges for security in MANETs. Traditional cryptographic mechanisms such as authentication and encryption are not capable of handling some kinds of attacks such as packet dropping by malicious nodes in MANETs. This paper presents a mechanism for detecting malicious packet dropping attacks in MANETs. The mechanism is depends on a trust module on each node, which is based on the reputation value computed for that node by its neighbors. The reputation value of a node is computed based on its packet forwarding behavior in the network. The reputation information is gathered, stored and exchanged between the nodes, and computed under different scenario. The proposed protocol has been simulated in a network simulator. The simulation results show the efficiency of its performance.

## 1 Introduction

Although the security objectives of both ad hoc networks and traditional networks are considered the same such as availability, confidentiality, integrity, authentication, and non-repudiation, the security issues involved in ad hoc networks are quite different due to their mobile and ad hoc constraints, i.e., limited computing and communication resources, dynamic network topology as well as the mobility of the nodes. In traditional networks, most trust evidences are generated via potentially lengthy assurance processes, distributed off-line, and assumed to be valid on a long term. In contrary, few of these characteristics of trust relations and trust evidences are prevalent in MANETs. Since the security solutions developed for the wired networks are not fit for scenarios, new security solutions become essential. Cryptographic primitives such as authentication and key distribution are the usual mechanisms used for implementing security in MANETs. However, these schemes cannot provide security against some attacks such as packet dropping attack by malicious nodes.

There are several approaches for security in MANETs [1][2]. The significant efforts done so far are mainly in the adaptation from the existing distributed trust model to ad hoc trust model. One approach of establishing trust among the nodes in a MANETs is by detecting misbehaving nodes that maliciously drop packets. These malicious nodes can be detected by utilizing the concept of *reputation*. The reputation of a node refers to the perception that another node has about its intention and activities. Reputation is a tool for motivating cooperation among nodes and so as to ensure that most of them exhibit good behavior in their activities. Each node in a network is assigned a reputation value as computed jointly by its neighbors. The higher the reputation value of a node the more trustworthy that node is. In this paper, a reputation-based distributed trust management scheme for MANETs is proposed. The nodes in a MANET collaborate to compute the reputation values of their neighbors, and identify the nodes for which reputation values fall below a pre-defined threshold value. The nodes having their reputation values below the threshold are identified as malicious.

The rest of the paper is organized as follows. Section 2 discusses some of the existing trust- and reputation-based schemes for MANETs. Section 3 describes the details of the proposed trust mechanism. Section 4 presents the results of simulation conducted on the proposed protocol. Section 5 discusses some future scope of work and concludes the paper.

## 2 Related Work

Different approaches exist for defining trust. Trust, in general, is a directional relationship between two entities and plays a major role in building a relationship between nodes in a network. Even though trust has been formalized as a computational model, it still means different things for different research communities such as public key authentication [3], electronic commerce [4], and P2P networks [5]. The *reputation* of an entity, on the other hand, has been defined as an expectation of its behavior based on other entities' observations or information about the entity's past behavior within a specific context at a given time [7]. In case of a MANET, the reputation of a node refers to how good the node is in terms of its contribution to routing activities in the network.

The *resurrecting duckling* security protocol proposed by Stajano et al. is particularly suited for devices without display and embedded devices that are too weak for public-key operations [8]. Eshenauer et al. have proposed a trust establishment mechanism for MANETs, in which a node in the network can generate trust evidence about any other node [9].

Among the more recent works, Repantis et al. have proposed a decentralized trust management middleware for ad hoc, peer-to-peer networks based on reputation of the nodes [10]. In the trust-based data management scheme proposed by Patwardhan et al., mobile nodes access distributed information, storage and sensory resources available in pervasive computing environment [11]. Sun et al. have presented a framework to quantitatively measure trust, model trust propagation, and defend trust evaluation system against malicious attacks [12]. Chang et al have proposed a trust-based scheme for multicast communication in a MANET [13]. Sen et al. have proposed a self-organized trust establishment scheme for nodes in a large-scale MANET in which a trust initiator is introduced during the network bootstrapping phase [14]. The authors have also proposed a distributed trust-based intrusion detection system for MANETs based on local observation and cooperation among nodes [15].

*Cooperation Of Nodes-Fairness In Dynamic Ad-hoc NeTworks* (CONFIDANT) is a security model for MANETs based on selective altruism and utilitarianism proposed by Buchegger et al. [6]. It is a distributed, symmetric reputation model that uses both first-hand and second-hand information for computation of reputation values. The proposed protocol in this paper has many similarities with the CONFIDANT protocol. However, the metrics for computing the reputation of a node in the proposed protocol are different from those used in CONFIDANT. The proposed protocol takes into account the historical data of the reputation of the nodes which makes the computed reputation values more robust. In contrast to the approach followed in CONFIDANT, the proposed mechanism broadcasts the reputation information to all neighbors of a node thereby making the protocol more reliable and fault-tolerant and hence more secure.

## 3   Trust Manager

Establishment of trust in a MANET requires successful detection of intruders and isolating them promptly so that they may not exploit any network resources. However, if one relies only on self-detecting misbehaviors, one may arrive at a wrong evaluation of trust. In fact, a node that is actually not sending any packets currently cannot detect selfish nodes in its neighborhood. As a consequence, collaboration between neighboring nodes becomes mandatory. In the proposed scheme, every node in the network monitors the behavior of its neighbors, and upon detecting any abnormal action from any of them, it broadcasts this information to other nodes in order to make them aware about its observation. The *neighbors* of a node *A* refer to all the nodes in the network those are one-hop distant from the node *A*.

The proposed mechanism builds trust through an entity, called the *trust manager* that runs on each node in the ad hoc network (Fig. 1). The Trust Manager has two main components: (i) *monitoring module* and (ii) *reputation handling module*.

### 3.1   The Monitoring Module

Each node in the MANET independently monitors the packet forwarding activities of its neighbors. This monitoring is related to the proportion of correctly forwarded packets with respect to the total number of packets to be forwarded during a fixed time window. Based on these statistics, if an anomaly is detected, the monitor informs the *reputation manager*, which analyses the packet loss information and take appropriate action. This is explained in Section 3.2

## 3.2 The Reputation Handling Module

The main functionality of the *trust manager* is the reputation information management. This functionality involves four major activities: (i) reputation information collection, (ii) reputation information formatting, (iii) reputation information maintenance, and (iv) reputation information rating. Each of these functions is described in detail in the following sub-sections.

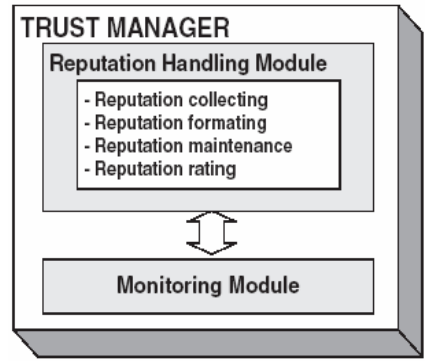

**Fig. 1.** The architecture of Trust Manager

### 3.2.1 Reputation information collection

This activity is the first step in the reputation information management process. In the proposed mechanism, the reputation information is collected in two ways: (a) sensing or direct monitoring, and (b) recommendations and accusations. In sensing or direct monitoring, a node *A* senses by itself misbehavior of one of its neighbor node say, *B* through its monitor module. In recommendations and accusations, the node *A* receives perceptions regarding a *presumed* misbehavior of a node *B* from its neighboring nodes. This process can be done in two ways: (i) *on-demand technique*, and (ii) *proactive technique*. In case of on-demand technique, a node *A* willing to compute a node *B*'s reputation, broadcasts a reputation request to its (i.e., node *A*'s) neighbors, and waits for reputation replies. Upon receiving the replies, it combines them in the way discussed in Section 3.2.4. In case of proactive technique, if a node *A* detects any misbehavior of a node *B*, it broadcasts the corresponding information to its neighbors even when it did not receive any reputation request related to node *B* from its neighborhood. The proactive technique is more suitable for protecting networks as the misbehaving information is broadcast as soon as an intrusion is detected. However, it has a high communication overhead on the network. The on-demand technique, on the other hand, has less overhead of communication. However, its security scope is limited, as all the nodes in the network do not have relevant security information all the time. For gathering reputation information, the proposed mechanism uses both the proactive and the on-demand techniques depending on the situation. To be more specific, the reputation information is exchanged as follows:

Let's first assume that two nodes *A* and *B* are present in a MANET and the node *A* detects that *B* is dropping packets. However, *A* will not broadcast this information to its neighbors unless *B*'s packet dropping rate crosses a pre-defined threshold value. Packet dropping by node *B* may be due to some physical problems of the node or because of its malicious behavior. Use of a threshold in monitoring packet loss ensures that the number of packets being dropped is high enough, and some new routes need to be computed regardless of the reason for the packet drop.

Let us now assume that a new node *D* has entered into the transmission range of node *A*. In this case, node *A* sends a reputation request about node *B* to the node *D* asking for its recommendations. The node *A* then combines the reputation reply that it receives from *D* with his own observation. This combined trust metric of the node *B* will reflect its final trust-worthiness. The method of computing the final reputation value is discussed in Section 3.2.4.

### 3.2.2 Reputation information formatting

The neighboring nodes in the MANET need to exchange the reputation information among each other. For exchanging reputation information, the proposed mechanism uses REP_MESS messages. A REP_MESS is an IP datagram with a *reputation header* inserted between the IP Header and the data payload. The reputation header consists of three fields: (i) REP_MESS_TYPE, (ii) NODE_ID, and (iii) REP_VAL.

REP_MESS_TYPE are of three categories: (i) REP_REQUEST, (ii) REP_RESPONSE, and (iii) REP_BROADCAST. The REP_REQUEST message type is used by a node when it requests for recommendations from its neighbors about a new node that is willing to join the network. The REP_RESPONSE message type is used by a node for replying to a recommendation request. For example, let us assume that a node *A* has already some reputation information regarding a node *B*. If node *A* receives a recommendation request regarding node *B*, it needs to generate a REP_MESS message with the REP_RESPONSE type and send it to the requester. The REP_BROADCAST message type is used when a node needs to broadcast some reputation information.

NODE_ID: it represents the IP address of the malicious node or the new node that is willing to join the network.

REP_VAL: it is the reputation value that the node detecting the misbehavior has computed, and stored in its reputation table.

### 3.2.3 Reputation information maintenance

The reputation information is evaluated at each node before it is locally stored or broadcast to its neighbors. The method of evaluation of reputation is discussed in Section 3.2.4.

Each node maintains a reputation table for storing reputation information for each of its neighbors. The node gets this information by either direct monitoring or through broadcast message received from some of its neighboring nodes. The reputation table for a node is updated whenever there is any change in its reputation value. The reputation table has two fields: NODE_ID and REP_VAL. The significances of these fields have been discussed in Section 3.2.2.

### 3.2.4 Reputation rating

Reputation value for a node is a number that can take values between 0 and 1. At the bootstrapping phase of the system, every node has reputation value 1. The reputation value for a node decreases if it exhibits any misbehavior. The reputation computation is done by taking into account the ratio of the correctly forwarded packets to the total number of packets that a node should forward in a given time window. The node *A* computes the reputation *r(A, B)* of node *B* using (1).

$$r(a,B) = \frac{\#ofpacketsforwarded}{\#ofpacketssent} \quad (1)$$

Let us now consider a MANET that consists of three nodes *A*, *B*, and *C*. Three different scenarios can be thought of as discussed below:

1. *Reputation computing during network establishment*: in this case, the nodes *A*, *B* and *C* meet for the first time. Each node creates an entry for the other two and assigns them reputation value of 1 to start with.

2. *Combining previous and current reputation values*: If node *A* detects misbehavior of node *B*, it needs to combine the new reputation value of node B with its previously stored reputation value. The combined reputation value is computed using (2).

$$r(A,B) = (1-\alpha)*r_{reptab}(A,B) + \alpha*r_{current}(A,B) \quad (2)$$

The factor *a* can take a value between 0 to 1. The value *r(A,B)* is a weighted sum of two components. The first part describes the node *B*'s reputation value already present in the node *A*'s reputation table. If node *A* have not met node *B* before, then the value $r_{reptab}(A,B)$ is set to 1 as mentioned earlier. The second part reflects contribution of node *B*'s new reputation value. As a node's previous reputations are also considered, the evaluation will be more consistent and seamless. Indeed, a good node that might have met some physical problems for a short time will not be punished and discarded as its reputation will surely increase again if it is relied on for forwarding data packets. Moreover, a node's reputation should seamlessly vary. If the reputation of a node fluctuates too fast, new routing paths will be frequently invoked, and the node's power will be quickly exhausted. However, the most recent reputation values are given higher weights by assigning higher values to the factor *α* (say 0.8 for example) in (2).

3. *Computing reputation when exchanging reputation information within a neighborhood*: According to the reputation collecting mechanism described in section 3.2.1, two cases need to be discussed in this scenario: (i) proactive and (ii) on-demand. These cases are presented below:

(i) *Proactive scenario*: in this case, a node *A* broadcasts reputation information regarding a node *B* to its neighborhood, in particular node to *C*, as soon as node *B*'s misbehavior is detected. In this case, node *C* computes the reputation of node *B* using (3) as follows:

$$r(C,B) = r_{reptab}(C,A)*r_{broad}(A,B) + (1-r_{reptab}(C,A))*r_{reptab}(C,B) \quad (3)$$

According to (3), the reputation of node *B* as maintained in node *C* is a weighted sum taking into account the proper perception of node *C* regarding node *B* and the perception of the broadcaster. The term $r_{broad}(A, B)$ is the reputation value of node B that the node A has broadcasted. The term $r_{reptab}(C, A)$ is the node *A*'s reputation value currently stored in node *C*'s reputation table. The new reputation value of the node *B* in the node *C*'s reputation table is computed based on the perception of the broadcasting node *A*. The relative weights of these two perception components will depend on the trustworthiness of the node *A*. If the node *A* is trustful, the value of $r_{broad}(A, B)$ is close to 1.

(ii) *On-demand scenario*: let us assume in this case that the MANET has *n* number of nodes denoted as: *A*, $N_1$, $N_2$, …..$N_{n-1}$. Let us imagine now that a new node *D* wants to join the network. The node *A* detects the presence of node *D*, and broadcasts a reputation request for node *D* in its neighborhood. If the node A receives responses from *p* number of neighbors, then node A computes the reputation value of node *D* using (4) as follows:

$$r(A,D) = \frac{\alpha * r_{reptab}(A,B)}{\alpha + \sum_{i=1}^{P} r_{reptab}(A,N_i)} + \frac{\sum_{i=1}^{P} r_{reptab}(A,N_i) * r_{reptab}(N_i,D)}{\alpha + \sum_{i=1}^{P} r_{reptab}(A,N_i)} \quad (4)$$

In this case, the all responses from different neighbors of node *A* are not treated equally. Node *A* treats each response based on the current reputation value the sender as existing in the reputation table maintained in it. Higher the value of *p*, i.e., more the number of neighboring nodes of *A* participating in the reputation value computation of node *D*, more accurate will be the finally computed reputation value. In the worst case, *p* =1, i.e., only one node from the neighbors of node *A* sends the response. In this case, the computed reputation value of node *D* will be least reliable. In this case, (4) will be effectively reduced to (3) with a multiplying factor.

## 4 Simulation

To test the performance of our mechanism, the 802.11 MAC layer implemented in network simulator *ns2* is used for simulation. The chosen parameters for simulation are presented in Table 1. Each node in the network is assumed to have a buffer with a capacity of 64 packets with FIFO interface queue. In the simulation, we have considered only the *Dynamic Source Routing* (DSR) protocol. However, our proposed trust model is applicable to any routing protocol for ad hoc networks.

Malicious nodes are simulated using a two-phase *Markov chain machine*. While in the *good* phase, the nodes do not drop any packets, in the *bad* phase, packets are dropped by malicious nodes based on a function. This function generates a random number between a maximum value (MAX_RATE) and a minimum value

(MIN_RATE). The Markov chain machine oscillates between both phases during a period of time ($t_{trans}$), which may be kept fixed or varied randomly. During the simulation, $t_{trans}$ is varied randomly between 100 sec to 200 sec. The traffic is simulated in the network by allowing 5 nodes to generate packets at the rate of 4 packets per sec.

**Table 1.** Simulation Parameters

| Parameters | Values |
| --- | --- |
| Simulation duration | 450 seconds |
| Simulation area | 1000 m * 500 m |
| Number of mobile nodes | 22 |
| Transmission range | 250 m |
| Movement model | Random waypoint |
| Maximum speed | 10 m/sec |
| Traffic type | CBR (UDP) |
| Number of malicious nodes | 5 |
| Host pause time | 300 sec |
| Max packet dropping rate | 8 packets/sec |
| Min. packet dropping rate | 1 packet/sec |

Fig. 2 shows the packet-dropping pattern by a malicious node over entire period of simulation. The *good* phase of the node is during the interval 140-200 sec and 400-450 sec. On the other hand, during the intervals 0-140 sec and 240-400 sec, the node is in the *bad* phase and drops packets varying from 5 to 30. Fig. 3 depicts the performance of the protocol in computing the reputation of a malicious node as performed by one of its neighbors. Here, both the nodes are on the routing path between the source and the destination. It is noted that the reputation value decreases from 1 to 0.25 in the first 140 sec, when the malicious node is in its *bad* phase as seen in Fig. 2. For the next 100 sec (140-240), the reputation of the node increases because the node is in its *good* phase. The reason for the slow rate of growth of reputation during the *good* phase of the node may be attributed to the packet loss due to temporary interference of the channel disturbance.

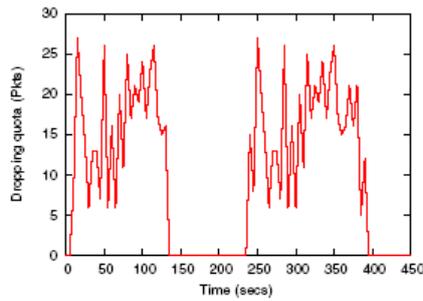 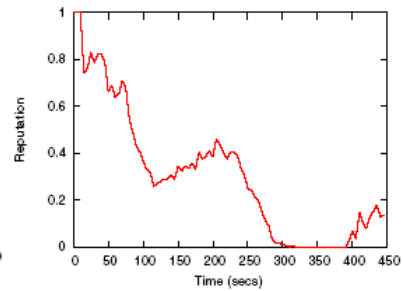

**Fig. 2.** Packet dropping by a malicious node    **Fig. 3.** Reputation of a malicious node

To study the performance of the protocol in a more realistic scenario, the value of $t_{trans}$ is varied between 0 and 5 sec. Two different situations were investigated: the reputation variation of a node in a low packet dropping scenario (Fig. 4), where

MAX_RATE and MIN_RATE are 5 and 0 packets respectively, and the reputation variation in presence of a high dropping configuration (Fig. 5), where MAX_RATE and MIN_RATE are 15 and 3 packets respectively. The reputation value oscillates between 0.58 and 0.78 in Fig. 4. This is because some packets are forwarded to destination; but some others are dropped during the *bad* phase.

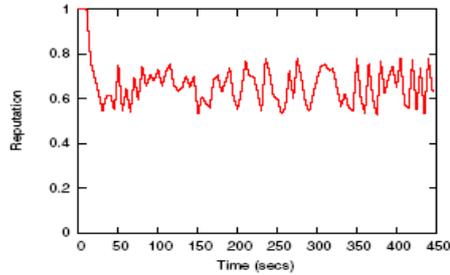 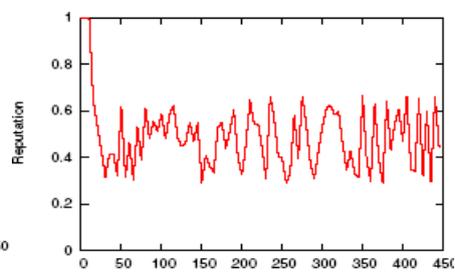

**Fig. 4.** Reputation in low packet drop rate  **Fig. 5.** Reputation in high packet drop rate

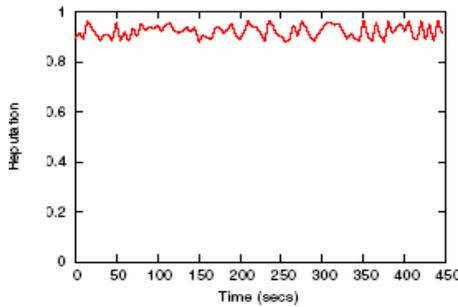 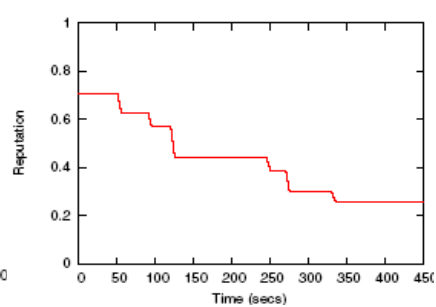

**Fig. 6.** Reputation of a non-malicious node  **Fig. 7.** Reputation of an accused node

Fig. 6 represents the pattern of variation of the reputation of a non-malicious node. It is observed that the highest and the lowest reputation value of the node are 0.98 and 0.85 respectively, which are quite reasonable values under the experimental setup. Finally, Fig. 7 illustrates the reputation variation of an accused node (say, *Y*) as computed by another node (say, *X*). Node *X* updates the reputation of *Y* by taking into account the trustworthiness of the broadcaster, and the previous reputation value of *Y*. As can be seen in Fig. 7, the reputation of *Y* started falling from its initial value of 0.7 after accusations were received.

## 5  Conclusion

In this paper, we have proposed a scheme that enables routing protocols in MANETs to detect malicious packet dropping by any node in the network. In the proposed scheme, each node in the network independently monitors the behavior of its neighbors and computes the reputation value for each of its neighboring nodes. Based on the reputation value of a node, its trustworthiness is determined. If the reputation of a node falls below a threshold value, it is no longer considered trustworthy by its neighbors and is identified as a malicious node. The results of simulation on the scheme show that it is quite effective in identifying malicious nodes in a MANET. Designing an efficient routing algorithm on top of this selfish node detection algorithm constitute a future work.